\documentclass[conference]{IEEEtran}
\ifCLASSINFOpdf
	\usepackage{graphicx}
	\graphicspath{{C:\Users\Minsang\Desktop\paper}}
	\DeclareGraphicsExtensions{.jpeg,.pdf,.png,.jpg}
	
	\usepackage{caption}
	\usepackage{subfigure}
	\usepackage[square,numbers,sort&compress]{natbib}
	\usepackage{setspace}
\else
\fi

\newcounter{myromancnt}
\renewcommand\themyromancnt{\Roman{myromancnt}}
\newcommand\myroman[1]{\setcounter{myromancnt}{#1}\themyromancnt}

\usepackage{algorithm}
\usepackage{algorithmic}
	
\begin{document}
%
\title{Requests Prediction in Cloud with a Cyclic Window Learning Algorithm}

\author{\IEEEauthorblockN{Min Sang Yoon$^a$,	Ahmed E. Kamal$^b$, and Zhengyuan Zhu$^c$}
\IEEEauthorblockA{$^a,^b$Department of Electrical and Computer Engineering\\$^c$Department of Statistics\\Iowa State University, IA, 50010\\ \{my222$^a$, kamal$^b$,zhuz$^c$\}@iastate.edu }
}

\maketitle

\begin{abstract}
Automatic resource scaling is one advantage of Cloud systems. Cloud systems are able to scale the number of physical machines depending on user requests. Therefore, accurate request prediction brings a great improvement in Cloud systems' performance. If we can make accurate requests prediction, the appropriate number of physical machines that can accommodate predicted amount of requests can be activated and Cloud systems will save more energy by preventing excessive activation of physical machines. Also, Cloud systems can implement advanced load distribution with accurate requests prediction. We propose an algorithm that predicts a probability distribution parameters of requests for each time interval. Maximum Likelihood Estimation (MLE) and Local Linear Regression (LLR) are used to implement this algorithm. An evaluation of the proposed algorithm is performed with the Google cluster-trace data. The prediction is implemented about the number of task arrivals, CPU requests, and memory requests. Then the accuracy of prediction is measured with Mean Absolute Percentage Error (MAPE). 

\end{abstract}


%
\IEEEpeerreviewmaketitle

\section{Introduction}
Predicting users' requests has been playing an important role in operating network systems because the accurate prediction can prevent systems from wasting operational cost and optimize resource utilization. Resource utilization is an especially important issue in Cloud systems. One advantage of Cloud systems is automatic scaling and management. Therefore, the accurate prediction of the requests in the Cloud systems will bring improvement in operating systems with more advanced resource management techniques. 

Cloud systems consists of a lot of computing and network devices. Thus, activating the appropriate number of devices based on the requests is an important issue in terms of saving energy. Data centers consumed 91 billion kilowatt-hours of electricity in 2013 and 50 percent of that is wasted due to lack of awareness of the traffic according to [1]. This data shows how the accurate prediction of the requests will be increasingly important in the future with an indisputable increase in energy consumption of data centers. Also, precise requests prediction has an effect on the performance of Cloud systems. If we inactivate overfull computing or network devices for energy saving purpose, it will cause a bottleneck and delay in handling requests.

There are many research that propose predicting an expected value of the requests with several methods. However, the prediction of the quantified number of requests is not the proper strategy for a stable operation of Cloud systems because user requests have heavy fluctuation frequently. Thus, we propose a strategy to predict parameters of the probability distribution of the requests in every regular interval instead of predicting the actual amount of requests. An advantage of predicting the probability distribution is that we can make the more flexible prediction. There are always request bursts in networks. If the system predicts a certain amount of requests based on the average of past requests, Cloud systems cannot prepare immediate variations of requests. However, if we are aware of the probability distribution of the requests during a given period, we are able to prevent request bursts by predicting the requests which corresponds to a high probability in the predicted probability distribution. 

For the prediction, we collect the history requests data of the Cloud and observe the histogram of the data. Based on the histogram analysis of the data, we decide the probability distribution model fitted to the collected data. Then, we induce  parameters of the probability distribution model with Maximum Likelihood Estimation (MLE) and save the parameter data to the prediction dataset. After we accumulate enough data for the prediction, the prediction model estimates parameters of the probability distribution with Local Linear Regression (LLR) by using a cyclic window approach. Parameters of the probability distribution are the time-dependent data, which means the parameters are changed as time goes by. So the parameters have obvious patterns during every interval. It will reveal the same pattern during a certain period in every day or every week. So our prediction model will make the prediction by using the accumulated dataset at every same time point through several periods. In order to reflect a change in trend of the requests, the prediction model maintains the dataset by replacing an old parameter data with the recent data.   

We use Google cluster-trace data to test our prediction model, which is real measurements of usage requests in Google cluster.  One week data is employed for a training data and the following week data is used for testing in order to verify whether the prediction model is accurate. We selected random time point to start collecting data so that the prediction model can be tested in more practical environment. There are three types of data for prediction, the number of task arrivals, CPU requests, and memory requests. Since various user requests are predicted and have a large variance in its values, we use Mean Absolute Percentage Error (MAPE) to normalize the error rate and assess the prediction accuracy. 

The rest of the paper is organized as follows. Previous works are reviewed in Section \myroman{2}. We will introduce mathematical methods and the prediction algorithm in Section \myroman{3}. A system model for an experiment will be demonstrated in Section \myroman{4} and the experiment results will be shown in Section \myroman{5}. The conclusions are given in Section \myroman{6}.

\section{Related work}
Requests prediction in Cloud system is accomplished by many researchers for the automatic scaling of systems. 

Akindele A. Bankole \emph{et~al}. employ machine learning technique for a predictive resource provisioning in Cloud [10]. Their prediction model is achieved with machine learning techniques: Neural Network, Linear Regression, and Support Vector Machine. They predict the CPU utilization, response time, and throughput based on collected data from virtual machines web servers and database servers. The prediction model generates prediction values in every given minute with machine learning techniques and measure error rate with MAPE and Root Mean Squared Error (RMSE). However, the prediction model did not show the high prediction accuracy. Their results show $24\%$ prediction error at a certain point of time in predicting the CPU utilization and a $21\%$ error in the response time prediction.

Sedeka Islam \emph{et~al}. present more advanced machine learning techniques for predicting resource usage [5]. The error correction Neural Network (ECNN) and the Linear Regression techniques are employed for prediction. They included a sliding window method to reflect the current state of the system. They have generated prediction values based on window sizes and evaluated them with MAPE, PRED(25), RMSE, and $R^2$prediction accuracy. The CPU utilization data is collected from the Amazon EC2 cloud through the TPC-W benchmark and prediction values are generated with the ECNN and Linear Regression method. The prediction values of CPU utilization has around $19\%$ error rate without the sliding window and has minimum $18.6\%$ error rate when they employ the sliding window.

Many statistical approaches are also applied to the prediction in Cloud. Bruno Lopes Dalmazo \emph{et~al}. propose the traffic prediction approach based on the statistical model where observations are weighted with Poisson distribution inside a moving window [9]. They consider the past information by means of a sliding window of size $\lambda$ and this window is applied by weighting the past observations according to the Poisson distribution with parameter $\lambda$. Dropbox trace data is employed for testing their prediction model and Normalized Mean Square Error (NMSE) evaluation method is utilized for the error measurement. The prediction model could achieve NMSE values between 0.044 and 0.112. The prediction is ideal when NMSE value is equal to zero and worse when it is greater than one. They could achieve the reasonably accurate prediction with this approach. 

They also suggest the traffic prediction model with a dynamic window approach [8]. The sliding window size is changed based on variance of the previous window size. The small variance indicates the predicted data is close to the meanwhile the high variance means the predicted data is spread out from the mean. So they update the window size in every prediction interval by considering the size of the variance in the previous prediction. The prediction accuracy is improved from $7.28\%$ to $495.51\%$ compared to the previous statistical model.

\section{Algorithm}
The prediction model estimates parameters of the probability distribution of future user requests in every predetermined period. We make the assumption that user requests have obvious patterns and the patterns are repeated periodically. For example, the request pattern of all Mondays will be a similar. Hence, the prediction model adopts the history parameter data at the same time point in order to make a prediction about the future. 

We introduce three time scales for the prediction periods: Pattern Period ($PP$), Target Period ($TP$) and Utilization Period ($UP$). The $PP$ is a cyclic interval that exhibits pattern repetition. The $TP$ is a unit duration for which we want to make a prediction. The $UP$ is a cyclic window that we use for predicting the activities in $TP$.  In the example, when we intend to predict the request distribution during a certain Monday by assuming the same pattern is repeated in every week. Then, we can set the $TP$ to a day and the $PP$ to a week because we assume the pattern of a day is repeated every week. If we only use the past Mondays$'$ data for the prediction, the $UP$ becomes one day. 

Since we assume patterns are repeated on the every $PP$, any time duration that we want to make a prediction corresponds to a certain $TP$ on the $PP$. Therefore, we can predict the request distribution during any time interval by corresponding them to a certain $TP$ on the $PP$. For the precise prediction, we accumulate data for several $PP$s. Although patterns of the traffic distributions will be similar on the same time point in every $PP$, the traffic amount will be different. In other words, we can say the distribution of the traffic shows similar form in every Monday, but we cannot ensure that the amounts of traffic will be same. Therefore, the prediction model can achieve higher prediction accuracy by accumulating data during several $PP$s. 

\begin{figure}[!h]
\centering 
\captionsetup{justification=centering}
\includegraphics[width=0.4\textwidth]{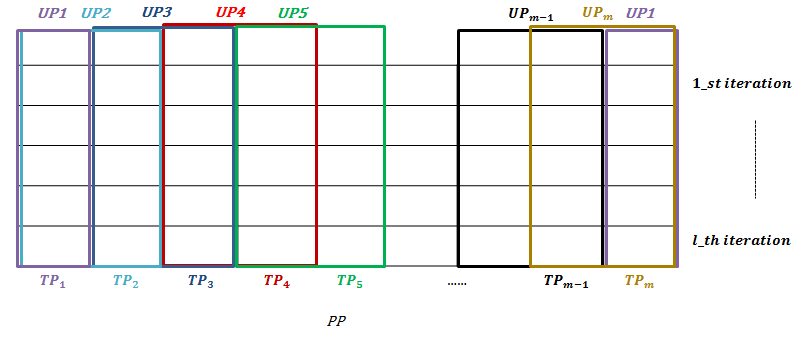}
\caption{Prediction dataset}
\label{fig:fig1}
\end{figure}

To implement prediction, the prediction dataset saves the past data in a $m\times l$ matrix. $m$ represents the number of $TP$s on the $PP$ and $l$ denotes the number of $PP$s we accumulate. In Figure 3, each vertical block corresponds to saved parameters of the probability distribution in each $TP$ during a $PP$. We start to stack the data from the first block of the first iteration. If we reach the $TP_m$'s block, which is the last TP, we move to the second iteration and stack the data from the first block of the second iteration, which means we have saved data during a $PP$. When the matrix is full, we go back to the first block of the first iteration and replace the old data to reflect the tendency of recent requests. 

Any time duration that we want to predict the traffic distribution for can be related to a certain $TP$ on the matrix. In order to make a prediction, we employ the $UP$ data. In Figure 3, we can see that distribution parameters of the $TP_m$ can be predicted by using the $UP_m$. We can set the size of the $UP$ depending on how many previous $TP$s will affect to the state of the current $TP$. For example, if we set the $UP$ to two days, we can say previous Sunday and Monday$'$s history parameters will affect to next Monday$'$s traffic distribution as we can see in Fig. 3. 

In this paper, we forecast the number of task arrivals during each target period. The first step of prediction, the prediction model constructs a histogram of user requests in every target period to observe distributions of requests. Then, it fits a probability distribution to the distribution of requests. When the probability distribution kernel is decided for fitting, we will adopt MLE in order to obtain parameters of the probability distribution in every observed period and the parameters are saved on the dataset. After the dataset accumulates enough data for prediction, it is able to predict parameters of a following target period. Local Linear Regression (LLR) will be employed to predict parameters of the future requests.

\subsection{Histogram}
A histogram is the graphical representation of the distribution of data. We can observe frequencies and overall distribution of given data through a graphical representation.

Histograms of requests are constructed in every regular interval to observe the distribution of requests. After observe the histograms, the prediction model decides which of the probability distribution model will be the closest to the actual distribution of requests. 

\subsection{Maximum Likelihood Estimation (MLE)}
MLE estimates parameters of a probability distribution when there are data corresponding to the probability distribution model. Our prediction model employ the Poisson distribution based on the histogram observation of the experiment data. Depending on accumulated request data, the prediction model will induce the Poisson distribution parameter by using MLE in every $TP$.

Poisson distribution has the only parameter $\lambda$. Since the prediction model has data through observation, the number of task arrivals, CPU, and memory request, it is able to induce parameter $\lambda$ by using MLE method. 
If we observe n independent datasets $X1$, $X2$, $X3$, $\ldots$, $Xn$ $iid$ Poisson random variables, maximum likelihood function L($\lambda$) will be:

\begin{equation}\label{eq:eq1}
L(\lambda)=\frac{\lambda^{X_1}e^{-\lambda}}{X_1\!}\frac{\lambda^{X_2}e^{-\lambda}}{X_2\!}\ldots\frac{\lambda^{X_n}e^{-\lambda}}{X_n\!}=\prod_{i=1}^{n}\frac{\lambda^{X_i}e^{-\lambda}}{X_i\!}
\end{equation}
If we take log in the equation, log likelihood function becomes:
\begin{eqnarray}\label{eq:eq2}
l(\lambda)=\sum_{i=1}^n(X_{i}\log \lambda-\lambda-\log X_i!)\nonumber\\
=log \lambda\sum_{i=1}^n X_i-n\lambda-\sum_{i=1}^n\log X_i!
\end{eqnarray}
We find maximum of $\lambda$ by finding the derivative of equation:
\begin{equation}\label{eq:eq3}
l'(\lambda)=\frac{1}{\lambda}\sum_{i=1}^n{X_i}-n=0
\end{equation}
, which implies the $\lambda$ that has closest distribution with observed histogram is:
\begin{equation}\label{eq:eq4}
\hat{\lambda}=\frac{\sum_{i=1}^n{X_i}}{n}=\bar{X}
\end{equation}

\subsection{Local Linear Regression (LLR)}
 LLR is one of the kernel smoother techniques for estimating a real value function, when no parametric model for this function is known. LLR combines much of the simplicity of linear least square regressions by fitting the line about the given $k$ number of points with the $N$ number of observed points. In the prediction model, the $k$ is corresponded to the number of $TP$s on $UP$s and $N$ is equivalent to the number of history parameters we will employ for the prediction, which is called to bandwidth. After fitting the line at every given point, the estimation functions $\hat{Y}(TP_k)$ are achieved as a value function with the $k$ numbers of values. 
$K_{h\lambda}(X_u,X_i)$ be a kernel defined by:

\begin{equation}\label{eq:eq15}
K_{h\lambda}(X_u,X_i)=D(\frac{\Vert X_i-X_u\Vert}{h_\lambda(X_u)})
\end{equation} 

The $D()$ is a positive real valued function in (5), which is decreasing when the distance between $X_i$ and $X_u$ increases. The $X_u$ is the given points and the $X_i$ is one of the observed data around $X_u$. Commonly used kernels include the Epanechnikov, biweigh and Gaussian function. 
For one dimension data, least-square method is employed for obtaining function value on the $X_u$. 

\begin{equation}\label{eq:eq16}
\min_{\alpha(X_u),\beta(X_u)}\sum_{i=1}^NK_{h\lambda}(X_u,X_i)\Big(Y(X_i)-\alpha(X_u)-\beta(X_u)X_i\Big)^2
\end{equation}

The $N$ is the number of history parameter near $X_u$, that we will employ in (6). Since we obtain parameters in each $TP_i$ by using MLE, the minimum of $\alpha(X_u)$ and $\beta(X_u)$ can be achieved by solving the weighted least square problem (6). 
If we assume the estimation function on $X_u$ is $\hat{Y}(X_u) = \alpha(X_u)+\beta(X_u)X_u$, the closed form solution of the estimation function is like:
\begin{equation}\label{eq:eq17}
\hat{Y}(X_u)=(1,X_u)(B^{T}W(X_u)B)^{-1}B^{T}W(X_u)y
\end{equation}
where:
\begin{equation}\label{eq:eq18}
y=(Y(X_1),\ldots,Y(X_N))^T
\end{equation}
\begin{equation}\label{eq:eq19}
W(X_0)=diag\Big(K_{h\lambda}(X_u,X_i)\Big)_{N\times N}
\end{equation}
\begin{equation}\label{eqLeq20}
B^T=\left( \begin{array}{cccc}1&1&\ldots&1\\X_1&X_2&\ldots&X_N \end{array}\right)
\end{equation}
By repeating this process about all given $k$ points, $X_u$, we can get real value estimation functions $\hat{Y}(X_u)$ about $k$ points.

\subsection{Cyclic Window Learning Algorithm}
We describe the algorithm that predicts parameters of the probability distribution of the future target periods by using a cyclic window approach. We assume the dataset has enough past data for the prediction and determined a probability distribution model for MLE. The algorithm will employ LLR to predict the probability distribution parameters of the future target periods and MLE for updating dataset. 

\begin{algorithm}
\caption{Cyclic Window Learning Algorithm}
\begin{algorithmic}[1]
\REQUIRE $PData_{m\times l}$,$TPdata_t$, $m$, $n$, and $l$
\ENSURE $PT_t$ and $AT_t$
\STATE $t=1$,$p=1$, and $w=1$
\WHILE {System operate}
\IF{$p<n$}
\STATE $UT_t=PData_{(m-n-p)\times l}:PData_{m\times l}$,\\$\qquad\quad PData_{1\times l}:PData_{p\times l},\forall l$
\ELSE
\STATE $UT_t=PData_{(p-n)\times l}:PData_{p\times l},\forall l$
\ENDIF
\STATE Implement LLR in terms of $UT_t$
\STATE $PT_t=\hat{Y}(UT_p)$ and select $k_{th}$ value
\STATE Update databased with actual parameter
\STATE $AT_p=MLE(TPdata_t)$
\STATE Update databased with $AT_p$
\STATE $PData(p, w)=AT_p$
\STATE $t=t+1$
\IF{$p<m$}
\STATE $p=p+1$
\ELSE
\STATE $p=1$
\IF{$w<l$}
\STATE $w=w+1$
\ELSE
\STATE $w=1$
\ENDIF
\ENDIF
\ENDWHILE
\end{algorithmic}
\end{algorithm}

The Algorithm 1 obtains the predicted parameters of target period ($PT$) for the prediction and actual parameters of target period ($AT$) to update dataset at every time period. The $m$ means the number of $TP$s included in a $UP$, which is equivalent to a window size. The $n$ is the number of $TP$s during a $PP$. The $l$ represents how many cycles of $PP$s will be stacked on the prediction dataset ($PData$). The $PData$ is the prediction dataset has the $m\times l$ dimension. The $TPdata_t$ means observed requests during an interval of the $TP$ at time $t$. The prediction model will obtain parameters of the $TP$ at time $t$ with MLE by using this data, $TPdata_t$. 

We initialize variables: $t$, $p$, and $w$ (line 1). The $t$ represents how many unit periods are passed after the algorithm starts and the $p$ is a corresponding position of the $TP$ about time $t$. Since we return to the initial position on the $PP$ when the $p$ reaches the end of the $PP$, the $p$ becomes a cyclic number from $1$ to $m$. The $w$ is a row position on the $m\times l$ $PData$. Too much data requires the complexity of the prediction and consumes too much time for the prediction. Therefore, we stack only the appropriate number of cycles on $PData$ by replacing the old data. The $w$ is which row position in the $PData$ will be updated.

First, the algorithm collects the data for the prediction from the $PData$ to the $UT_t$ (line 2 $-$ line 7). If the position $p$ is less than the window size $n$, we need to employ the data from the end of the $PData$ because the $p$ is cyclic. So we collect the data from $m-n-p_{th}$ to $m_{th}$ columns' data and from the first to $p_{th}$ columns of $PData$ (line 3 $-$ line 4). If the $p$ is greater than the window size $n$, we collect previous $n$ columns data from the point $p$ on the $PData$ (line 5 $-$ line 7). 

We implement LLR about the collected data, $UT_t$, and obtain the prediction value about time $t$ (line 8 $-$ line 9). In order to update the $PData$, we need to obtain the actual probability distribution parameters of observed requests. We apply MLE about the data $TPdata_t$ and update corresponding data block in $PData$ (line 10 $-$ line 13).
We update the position on the $PData$ for next prediction. We update the $t$ for predicting next time point (line 14). We just increase the $p$ if the position of the $p$ is still on $PP$. If the $p$ exceeds the size of $PP$, $m$, it goes back to the initial position of the $PP$. We also update the $w$ when $p$ goes back to the initial position because that means one row of $PData$ is filled with new data. The $w$ increases when the $p$ increases. However, the $w$ becomes one if the $w$ exceeds $l$, which is vertical size of the matrix $PData$ (line 15 $-$ line 23).

\section{System model}
\subsection{Google Cluster Data collection (Arriving Tasks, CPU, and Memory)}
The experiment is implemented with Google cluster-usage traces data [14]. Google cluster is a set of computing resources composed of thousands of machines. A job is composed of several tasks which can be processed separately. So each task will be a unit of a process. We consider the number of task arrivals, CPU requests, and Memory requests of tasks. Each task has a timestamp which represents when the task arrives at the cluster. Therefore, the distribution of the number of task arrivals can be observed by using the timestamp. The cluster data also contains the CPU and memory requests of each task. The CPU requests show core counts or core-seconds/second of tasks and the memory requests represent how much bytes each task requires. 
The cluster starts measurement 600 seconds after the system is operated and has accumulated data for one month approximately. We select a random point to collect data for the prediction model. One weak data is sampled as a training dataset for the prediction modeling and the following week data is employed to test the accuracy of the prediction model.

\section{Experiment}
\subsection{Patterns of data}
All experiments are conducted by Matlab. Basically, user requests have a regular pattern. Requests increases during the daytime and weekdays more than the nighttime and weekend. We could observe some patterns by analyzing the Google trace data. The start time of trace is randomly chosen in data. So we do not know what the exact date or time of measurement points. However, our assumption is arriving tasks will show regular patterns in every interval. 

\begin{figure}[h!]
\centering 
\captionsetup{justification=centering}
\includegraphics[width=0.4\textwidth]{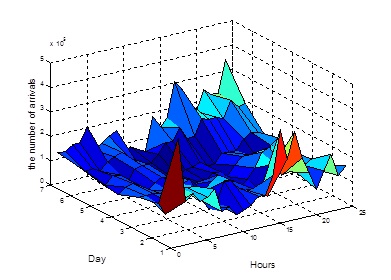}
\caption{Daily Pattern Analysis}
\label{fig:fig2}
\end{figure}

Figure 2 shows patterns of task arrivals during a week. The number of task arrivals is counted for every one hour, so we could observe how many tasks arrive at the cluster in every hour during the week. The hourly pattern shows high peaks for the first 4 hours and the last 10 hours. So we can assume daytime starts around the tenth hours from the measurement because the number of tasks is increased from the tenth hour and end at the fourth hour in Figure 2. If we see the daily pattern, the first two days and the last two days show higher requests.  In the same way, we can assume weekdays start on the fourth day from the measurement by assuming requests increases during weekdays rather than the weekend. 

The daily pattern analysis of task arrivals presents obvious patterns of the incoming requests. Therefore, the cyclic data collection should be an effective approach for the prediction.

The daily pattern analysis of the CPU and memory requests also have demonstrated the similar patterns with task arrivals. The CPU and memory requests have had high requests when the number of task arrivals increases and had low requests during free periods.  

According to our observation, we have decided to set the $PP$ to one week because we could observe hourly and daily patterns of requests repeatedly during every week. 

\subsection{Histogram analysis of data}
According to the experiment, we found that too short $TP$ is not enough to observe apparent patterns of distribution of requests. Thus, the $TP$ is set to 30 minutes based on empirical observation. The prediction model predicts distribution parameters in every 30 minutes.  

\begin{figure}[h!]
\centering 
\captionsetup{justification=centering}
\includegraphics[width=0.4\textwidth]{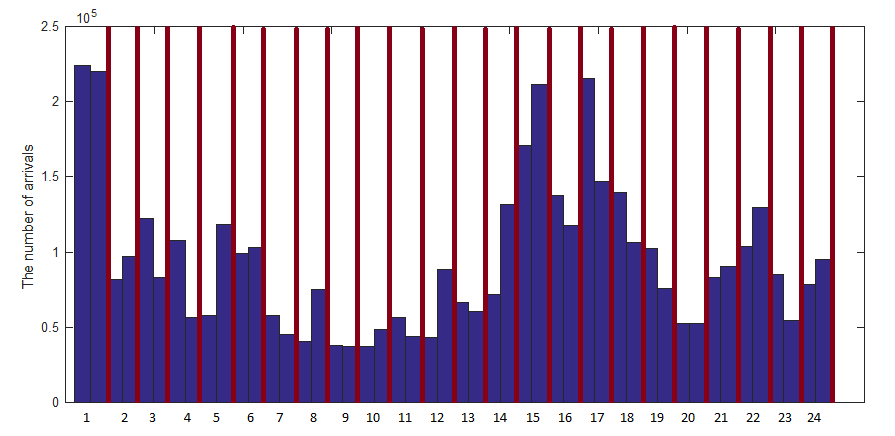}
\caption{The number of tasks during the day}
\label{fig:fig3}
\end{figure}

As patterns are observed in Figure 2, the number of task arrivals show an apparent pattern depending on the time. Figure 3 represents the number of task arrivals in every 30 minute during the first day. For the first 4 hours and the last 10 hours, Figure 3 presents high rate of incoming tasks. 

Histograms of each target period are different depending on the total number of task arrivals. Histograms have high peaks in the more right side during the busy hours and they have high peaks in the more left side during the free hours. 

\begin{figure}[h!]
\centering 
\captionsetup{justification=centering}
\includegraphics[width=0.4\textwidth]{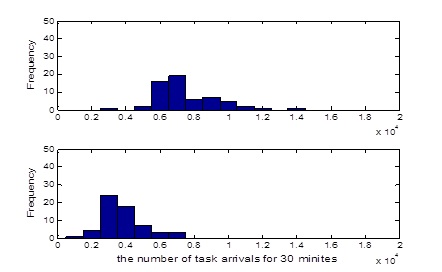}
\caption{Distribution of the first day}
\label{fig:fig4}
\end{figure}

Figure 4 is a histogram of the first hour. The first histogram exhibits the distribution of task arrivals during the first half hour and the second histogram presents the distribution of the second half hour. The first histogram has a peak in more right side than the second histogram because the cluster received higher task arrivals during the first half hour than the second half hour.

This trend is similar to the Poisson distribution. The Poisson distribution has a high peak in the more right side when the rate parameter $\lambda$ is high. Therefore, the observed data will be fitted to the Poisson distribution by using MLE to obtain parameters of the Poisson distribution in every duration. 

\subsection{Maximum Likelihood Estimation of distributions}
 
\begin{figure}[h!]
\centering 
\captionsetup{justification=centering}
\includegraphics[width=0.4\textwidth]{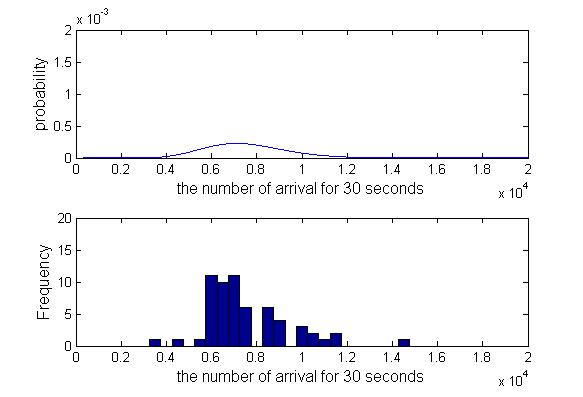}
\caption{MLE of the first hour}
\label{fig:fig6}
\end{figure}

 MLE is employed to obtain parameters of the Poisson distribution in every target period.
The estimated Poisson distribution are achieved about the first half hour histogram by using MLE in Figure 5. The first graph shows the Poisson distribution with estimated parameters and the second graph is a histogram of task arrivals during the first half hour. We can observe the estimated Poisson distribution has a similar distribution with the histogram of data. The prediction model implements MLE in every 30 minute and saves them to the prediction dataset. 

\subsection{Time Dependent Parameter Estimation}
Parameters are induced from the request data in every 30 minute. Parameters generated at the same time point on the $PP$ are stacked in the same column of the prediction dataset. Predicted parameters are achieved by implementing LLR about the corresponding $UP$. For example, if we implement LLR about the $UP$ including the first to the $10_{th}$ $TP$ to predict the duration corresponding to $10_{th}$ $TP$, the last point value of LLR function becomes the prediction value of the $10_{th}$ target period. Prediction values are changed depending on how to set the utilization periods and how much bandwidth we adopt for LLR. The bandwidth represents how many near data are included when we predict the function value of a certain point. 
 
\begin{figure}[h!]
\centering 
\captionsetup{justification=centering}
\includegraphics[width=0.4\textwidth]{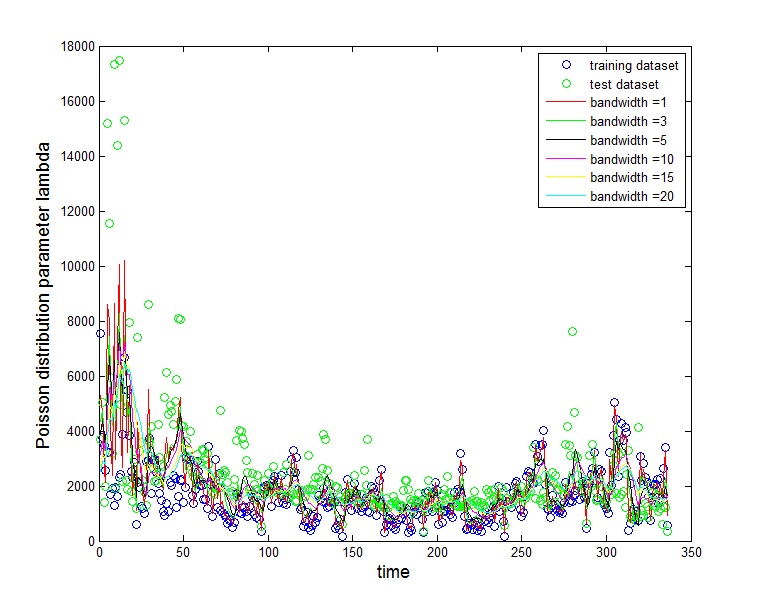}
\caption{Poisson distribution parameter $\lambda$ estimation of arrival tasks}
\label{fig:fig9}
\end{figure}

Figure 6 represents the prediction of Poisson distribution parameter $\lambda$ of arriving tasks during week. Since we set the $TP$ to 30 minutes, we have 336 target periods during the week. Blue points represent parameter values of training dataset in each $TP$ and green points are parameter values of test dataset in each $TP$. Solid lines represent parameter prediction values for different values of bandwidth. We set the $UP$ to 25 hours in Fig. 4, which means that prediction value is obtained base on the last $25_{th}$ hours data. Parameter $\lambda$ is equivalent to mean number of arrivals during 30 minutes. We can observe that the graph has a regularly repeated pattern. It has seven high peaks in the graph, which means similar patterns repeated during the week.

\begin{figure}[h]
\centering
\subfigure[]{
\includegraphics[width=.35\textwidth]{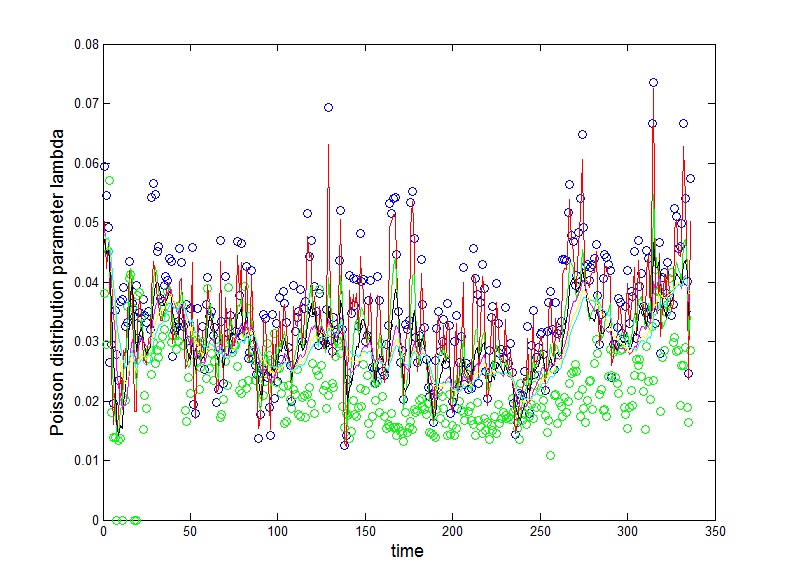}
\label{fig:nrGroup}
}
\subfigure[]{
\includegraphics[width=.35\textwidth]{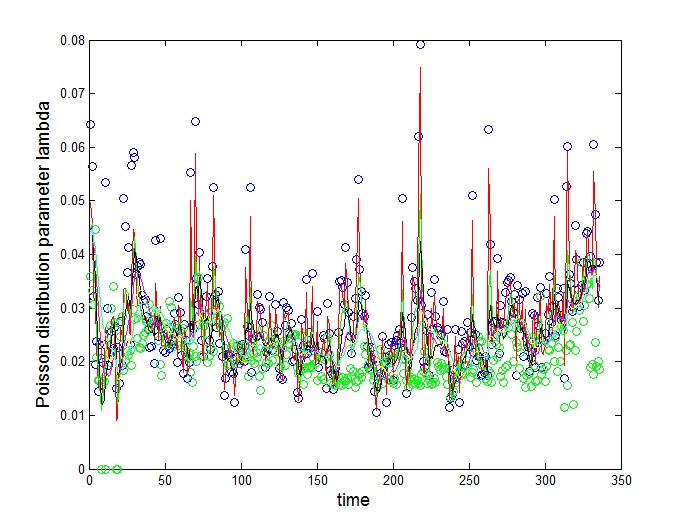}
\label{fig:overallResult}
}
\caption{Poisson distribution parameter $\lambda$ estimation \\(a)CPU (b)Memory}
\end{figure}

Prediction values of parameters are obtained about the CPU and memory requests as well in the same method. Figure 7 is the parameter prediction of the CPU and memory requests. We obtain prediction values in the same method with task arrivals. 

\subsection{Error assessment}
In order to quantify an accuracy of the prediction, we measure Mean Absolute Percentage Error (MAPE) between the prediction data and the test dataset. MAPE expresses an error rate as a percentage. So we can compare the prediction accuracy of task arrivals, CPU requests, and memory requests with a normalized error rate value.  

\begin{equation}
MAPE=\frac{1}{n}\sum_{j=1}^n \frac{\vert P_j-T_j\vert}{T_j}
\end{equation}

The $P_j$ is a predicted value of a target value, $T_j$. MAPE value is equal to zero when the prediction model is the perfect fit to the target value and increased when the prediction is not properly fit to target values. 

\begin{figure}[h]
\centering
\includegraphics[width=.4\textwidth]{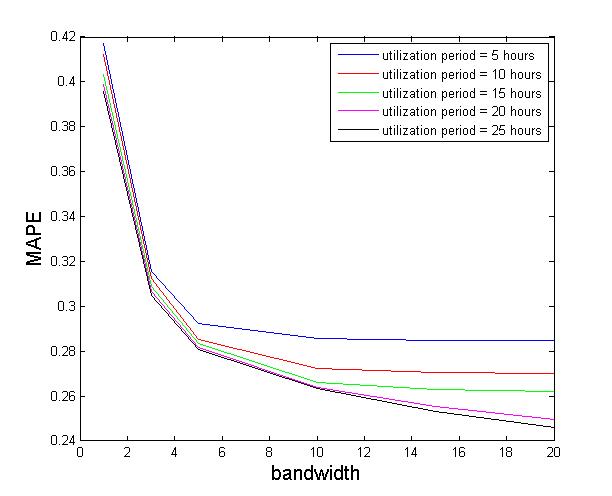}
\label{fig:nrGroup}
\caption{MAPE measurement of task arrivals prediction}
\end{figure}

Figure 8 is MAPE measurement graph of the Poisson distribution parameter $\lambda$. The parameter $\lambda$ has MAPE range between 0.3885 and 0.5194. If we consider the ideal state of MAPE is zero, the prediction model has enough prediction accuracy. The prediction model could achieve a higher accuracy with the longer $UP$, which is equivalent the window size because increasing the $UP$ means employing more previous data for the prediction. However, the large $UP$ requires more complexity of a computation and consumes more time. In other words, a proper selection of the $UP$ is required to satisfy both of the prediction accuracy and the computation time. Choosing the best bandwidth is also an important issue in order to reduce the prediction error. Too small bandwidth causes very spiky estimates while large bandwidth leads over smoothing. If data values are spread widely, the smaller bandwidth will not acquire the higher prediction accuracy. 

\begin{figure}[h]
\centering
\includegraphics[width=.4\textwidth]{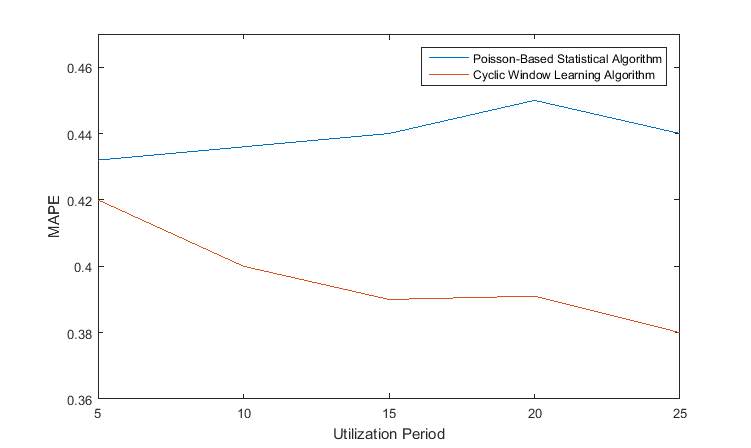}
\label{fig:nrGroup}
\caption{MAPE measurement of task arrivals prediction}
\end{figure}

We compare the prediction accuracy of cyclic window learning algorithm with Poisson based statistical algorithm that suggested in [9]. We selected the prediction values obtained with bandwidth 20 and compared them with the prediction result of Poisson based statistical algorithm in Figure 9. In the result, we could observe that the cyclic window learning algorithm presents slightly better prediction accuracy when we employed small $UP$s. However, the cyclic window learning algorithm could improve the performance up to 13.6$\%$ when we employ more $UP$s data. 

\section{Conclusions}
We propose a novel approach for the request prediction in Cloud systems. Instead of predicting an actual amount of requests, our prediction model estimates parameters of the probability distribution during the given period. We accumulate the historical data of the system and a cyclic window approach to utilize data with MLE and LLR. In the experiment with Google cluster-trace data, we could ensure advanced performance of the prediction algorithm. Our prediction model achieves the very low level of error rates in predicting the probability distribution parameters.

{
}
\end{document}